\documentstyle[12pt]{article} \topmargin=-0.4in

\begin{document} 
\begin{flushright}
UCTP-102-01
\end{flushright}
\begin{center} 
{\LARGE \bf Monopoles and vortices in pure gauge theories and in Higgs theories
}\\
\vskip0.1in
P. Suranyi\\
University of Cincinnati, Cincinnati, Ohio 45221-0011, USA
\end{center} 
\begin{abstract} Smeared Abelian and center gauges are introduced in pure non-Abelian lattice
gauge theories. Popular Abelian and center gauges are limits of smeared
gauges.  Smeared gauges are also shown to be equivalent to Higgs theories.  As a
result, distributions and interactions of monopoles and vortices, which are  objects responsible for
confinement in pure gauge theories, can be studied  by investigating classical solutions of Higgs
theories.
\end{abstract}
\section{Introduction}
Though there is no analytic proof of confinement in non-supersymmetric nonabelian gauge theories 
lattice simulations provide  ample numerical evidence that nonabelian gauge theories 
confine quarks and gluons~\cite{lattice}.  
Considerable effort has been spent to understand the
underlying physical mechanism behind confinement. The two most popular models, explaining the 
physical reasons behind confinement, are  the dual superconductor model~\cite{thooft1} and
the 
$Z_N$ vortex condensation
model~\cite{thooft3}~\cite{nielsen0}~\cite{mandelstam1}~\cite{mack}. 

Continuum gauge theories require gauge fixing.  On the lattice, gauge fixing is not required, but
possible. In the dual superconductor model the nonabelian part of the gauge is fixed, while a  Cartan
subgroup is not.  As gauge fixing leaves behind a theory with Abelian gauge symmetry, this is called
Abelian gauge fixing. The gauge fixing procedure has defects, which can be interpreted as
monopoles.  The gauge field, projected to its abelian components, and
monopoles form a Coulomb gas that confines charged objects~\cite{lattice}.

In the $Z_N$ magnetic vortex model of $SU(N)$ gauge theories the gauge is fixed completely,
except for the discrete center of the group, $Z_N$.  This is called center gauge fixing. The defects
of this gauge fixing are lines, representing magnetic fluxes, vortices, in three dimensional space.  
The condensation and percolation of vortices can be easily shown to lead to an area law of Wilson
loops, which is tantamount to confinement. 

 Lattice
studies of abelian projected theories~\cite{abelian1}~\cite{abelian2} found good agreement
between numerical values of gauge invariant quantities (string tension, chiral condensate, etc.) 
obtained with and without  abelian projection.  This was assumed to be an indication of the validity
of the dual superconductor picture.  

Center projected pure lattice gauge theories~\cite{vortex1}~\cite{vortex2}~\cite{vortex3} are
$Z_N$ gauge theories. $Z_N$ gauge theories  are theories of interacting vortices.  Averages of
Wilson loops and of other gauge invariant quantities calculated in the projected $Z_N$ gauge
theory  were also  in good agreement with numerical values of gauge invariant quantities  obtained 
without gauge fixing.  This agreement implies  that the center vortex picture  is  also a viable
candidate for explaining confinement. 

The analytic study  of monopoles and magnetic vortices in pure gauge theories
 is, however, extremely difficult.  Among existing studies,  Polyakov's 
classical
Coulomb gas model model~\cite{polyakov2},  and  recent works 
defining vortices in the continuum~\cite{analytic1} are of this nature.  Strong
coupling expansion ~\cite{analytic2} gives credence to the vortex condensation picture, as well. 
The analytic study of monopoles and vortices in gauge fixed pure gauge theories is very important. It
is conceivable that  finding an analytic proof of confinement may require an analytic proof of the
existence and condensation of these classical objects in pure gauge theories.  

Topological solutions, such as monopoles and vortices, appear in another class of theories under
more controlled circumstances.  
Monopoles were found in Higgs theories with a single adjoint
representation Higgs boson~\cite{thooft4}~\cite{polyakov} and
vortices were found in Higgs
theories with two or more
Higgs bosons~\cite{devega}~\cite{hasenfratz}~\cite{schwartz}~\cite{schaposnik}, as
topologically stable classical solutions. 

In an earlier
work we conjectured that  monopoles and vortices found in gauge fixed pure gauge theories 
are connected with monopoles and vortices appearing as classical solitons in Higgs
theories~\cite{suranyi}. In this paper we intend to investigate this  relationship and establish a firm
link among these objects.
Using  the link between gauge fixed pure gauge theories and of Higgs theories the properties of
monopoles and vortices and their interactions can be studied analytically  in Higgs theories and the
conclusions can be carried over to gauge fixed pure gauge theories.   

We will
rely on lattice realizations of gauge theories and Higgs theories throughout this paper.  We will
assume that at sufficiently small gauge coupling they are close to continuum theories.  For the
sake of simplicity we will only consider
$SU(2)$ gauge theories, but our conclusions can be generalized to other nonabelian gauge theories
without much difficulty. 

\section{Abelian gauge fixing and Higgs theories}

We will first consider Abelian gauge fixing procedures in pure lattice gauge theories.
The most popular gauges are the Maximal Abelian gauge (MAG) and the Laplacian Abelian gauge
(LAG). These are opposite limits of a continuous set of gauge fixing
procedures that we will define below.  All of these procedures {\em use lattice gauge configurations
generated without gauge fixing}.

MAG is
defined~\cite{abelian1} by maximizing a functional, $S_{\rm gf}^V$, of the gauge field over gauge
transformations, 
$V(\mbox{\boldmath$r$})$, 
\begin{eqnarray}
F_{\rm gf}(U^V)&=&\frac{1}{2}\sum_{\mbox{\boldmath$r$},\mu}{\rm Tr}\left[\sigma_3
U^V_\mu(\mbox{\boldmath$r$})\sigma_3 U_\mu^{V,\dagger} (\mbox{\boldmath$r$})\right],
\label{mag}
\end{eqnarray}
where
$U^V_\mu(\mbox{\boldmath$r$})=V(\mbox{\boldmath$r$})U_\mu(\mbox{\boldmath$r$})
V(\mbox{\boldmath$r$}+\mbox{\boldmath$
\mu$}
)$.  $\mbox{\boldmath$ \mu$}$ represents a
lattice vector in the direction of the $\mu$th axis.

For future purposes we rewrite $F_{\rm gf}(U^V)$ in terms of quantities defined in the adjoint
representation.  As the choice of gauge is arbitrary we will choose $U_\mu(\mbox{\boldmath$r$})$
as a gauge field at the maximum of $F_{\rm gf}(U^V)$, at
$V(\mbox{\boldmath$r$})=e^{i\gamma_3\sigma_3}$.  Then we write $S_{\rm gf}^V$  as
\begin{eqnarray}
F_{\rm gf}(U^V)&=&
\sum_{x,\mu} g^A_{3a}(\mbox{\boldmath$r$})U_\mu^{A,ab}(\mbox{\boldmath$r$})
g^A_{3b}(\mbox{\boldmath$r$}+\mbox{\boldmath$\mu$}),
\label{max-ab-gauge}
\end{eqnarray}
where we defined the adjoint representation
gauge transformation as
\begin{equation}
g^A_{ab}(\mbox{\boldmath$r$})=\frac{1}{2}{\rm Tr}\left[\sigma_a
V(\mbox{\boldmath$r$})\sigma_bV^\dagger(\mbox{\boldmath$r$})\right]
\label{g-adjoint}
\end{equation}
and the adjoint representation gauge field as 
\[
U_\mu^{A,ab}(\mbox{\boldmath$r$})=\frac{1}{2}{\rm Tr}\left[\sigma_a
U_\mu(\mbox{\boldmath$r$})\sigma_bU_\mu^\dagger(\mbox{\boldmath$r$})\right].
\]
$g^A_{ab}$ and $U_\mu(\mbox{\boldmath$r$})$ are orthogonal matrices.  Note that the Cartan
(Abelian) subgroup of the gauge group leaves the vector
$g^A_{a3}$ invariant. In $SU(2)$ the normalized vector $g_{a3}^A(\mbox{\boldmath$r$})$ fixes
only two of the parameters of the gauge group.

LAG relaxes the constraint $\sum_a[ g^A_{a3}(\mbox{\boldmath$r$})]^2=1$~\cite{abelian2}. It
maximizes (\ref{max-ab-gauge}) by finding the eigenvector of largest eigenvalue of the 
quadratic form defined  by the right hand side of (\ref{max-ab-gauge}).
The local normalization of the real vectors $g_{a3}(\mbox{\boldmath$r$})$  then determines the
gauge transformation through (\ref{g-adjoint}).  The conceptual advantage of LAG over MAG is
that the trajectories of monopoles can be defined as singularities of the gauge fixing procedure, in
agreement with the original ideas of `t Hooft~\cite{thooft1}.  At points where
$g_{a3}(\mbox{\boldmath$r$})$ vanishes the gauge transformation becomes singular. These points
define the world lines of monopoles in four dimensional space.

We introduce now a gauge, that interpolates between MAG and LAG, by defining a
functional~\cite{suranyi}
\begin{eqnarray}
Z_A(U^V)&=&\int_0^\infty df(\mbox{\boldmath$r$})\,f^2(\mbox{\boldmath$r$})\exp\left\{-
\tilde F_{\rm gf}(U^V,f)\right\},
\label{intermediate}
\end{eqnarray}
where $\tilde F_{\rm gf}(U^V,f)$ is defined as 
\begin{equation}
\tilde F_{\rm
gf}(U^V,f)=-\sum_{\mbox{\boldmath$r$}}\sum_\mu\left\{f(\mbox{\boldmath$r$})
f(\mbox{\boldmath$r$}+\mbox{\boldmath$\mu$})\frac{1}{2}{\rm
Tr}\left[\sigma_3U_\mu^V(\mbox{\boldmath$r$})\sigma_3U_\mu^{V\dagger}(\mbox{\boldmath$r$})\right]-f
^2(\mbox{\boldmath$r$})\right\}+\frac{\lambda}{\eta^2}
\sum_{\mbox{\boldmath$r$}}\left[f^2(\mbox{\boldmath$r$})-\eta^2\right]^2.
\label{higgs-action0}
\end{equation}
 
The prescription for finding the appropriate abelian gauge is {\em to find the maximum} of
$Z_A(U^V)$ in terms of the   $V(\mbox{\boldmath$r$})$  (or rather $g^A_{a3}$) and
$f(\mbox{\boldmath$r$})$.   The gauge defined by
 (\ref{intermediate}) is  similar to the LAG gauge-fixing functional, in the
sense that the optimal local vector
$f(\mbox{\boldmath$r$})$ vanishes at certain points (or rather, owing to  three constraints, on
curves in spacetime).  At these points the gauge transformation becomes singular.  Just like in LAG,
these points correspond to monopoles.

Notice now that the $\lambda\to\infty$ limit is equivalent to MAG.  Indeed,
this limit freezes out the integration over $f(\mbox{\boldmath$r$})$ and leaves one with the
exponential of  (\ref{max-ab-gauge}), multiplied by $\eta^2$.  Furthermore, the $\lambda\to0$ limit
is equivalent to LAG.  The introduction of the field
$f(\mbox{\boldmath$r$})$ is equivalent to relaxing the constraint on the magnitude of the  gauge
group element in adjoint representation, due to the replacement
$g^A_{3a}(\mbox{\boldmath$r$})\to f(\mbox{\boldmath$r$})g^A_{3a}(\mbox{\boldmath$r$})$.
Only the sum,
$\sum_{\mbox{\boldmath$r$}a}
f^2(\mbox{\boldmath$r$})(g^A_{0a})^2(\mbox{\boldmath$r$})=\sum_{\mbox{\boldmath$r$}}
f^2(\mbox{\boldmath$r$})$
is restricted, to make the integral finite. The general case of finite
$\lambda$ that interpolates between these two gauge fixing methods is also an appropriate Abelian
gauge and will henceforth be called the general Abelian gauge (GAG). 

Next we propose  an alternative to GAG, a
conventional gauge fixing term, $S_{\rm gf}$,  inspired by the functional
$Z_A(U^V)$ and defined as
\begin{equation}
e^{-S_{\rm gf}(U^V)}=Z_A(U^V)e^{-\tilde S},
\label{new-gauge-fix}
\end{equation}
where 
\begin{equation}
e^{\tilde S}=\int dV'(\mbox{\boldmath$r$}) Z_A(U^{V'}).
\label{stilde}
\end{equation}
The integral of (\ref{new-gauge-fix}) over all gauge
transformations is 1 and
the modified action, $S=S_G+S_{\rm gf}$,  leads to the {\em same expectation value of  all gauge
invariant quantities as the original plaquette action},
$S_G$. We call this a smeared abelian gauge (SAG). Unlike in GAG, gauge configurations are
understood to be generated in the presence of the gauge fixing term.  

It is comparatively easy to establish the relationship between SAG and GAG.  What is more
important this relationship extends to non-gauge invariant quantities, such as gauge field
configurations. Consider the Boltzmann factor  in SAG,
\[
e^{-S_G(U^V)-S_{\rm gf}(U^V)}.
\]
 $S_{\rm gf}(U^V)$ has a sharp
minimum at $g^A_{a3}(\mbox{\boldmath$r$})=\delta_{a3}$.   The integral in the denominator of
(\ref{new-gauge-fix}) can be evaluated by the saddle point method in the nonabelian parameters of
the gauge group.  Up to a determinant, it gives
$Z_A(U)$, the maximum of the numerator of (\ref{new-gauge-fix}), over
$V(\mbox{\boldmath$r$})$. Though the integral of the Boltzmann factor over
$U_\mu(\mbox{\boldmath$r$})$ is gauge invariant, $Z_A(U^V)$ will give an appreciable
contribution to the integral over
$U_\mu(\mbox{\boldmath$r$})$ only if $g^A_{a3}(\mbox{\boldmath$r$})\simeq \delta_{a3}$. 
Thus, when one generates an ensemble of gauge field configurations using the Boltzmann factor
$\exp\{-S_G-S_{\rm gf}\}$  one obtains only configurations that are very
close to GAG configurations. 
The
equivalence between GAG and SAG would become exact at
$\eta\to\infty$.\footnote{One can rescale function $f$ by $\eta$ and then it is easy to see that all
values of $f$ and $g^A$ but those maximizing the gauge fixing term are frozen out.} In other
words, at large $\eta$ SAG is only slightly smeared around GAG.  

We  now consider SAG from a different angle.  Aside from $\tilde S$, the action
 is just that of a traceless adjoint Higgs theory. To prove this statement, we use the gauge
invariance of the measure, $dU_\mu(\mbox{\boldmath$r$})$,  of $S_G$, and of $\tilde S$ to
introduce an extra integration over gauge transformation $V$, and write  the partition function as
\[
Z=\int dV(\mbox{\boldmath$r$})\int dU_\mu(\mbox{\boldmath$r$}) e^{-S_G(U)-S_{\rm
gf}(U^V)}=\int dU_\mu(\mbox{\boldmath$r$})
\int d\Phi(\mbox{\boldmath$r$})\, e^{-S_G(U)-S_{H}(U,\Phi)-\tilde S(U)},
\]
where we defined the Higgs action as
\begin{equation}
S_{H}(U,\Phi)= -\frac{1}{2}\sum_{\mbox{\boldmath$r$}\mu}{\rm
Tr}\left[\Phi(\mbox{\boldmath$r$})U_\mu(\mbox{\boldmath$r$})\Phi(\mbox{\boldmath$r$}
+\mbox{\boldmath$\mu$})U^\dagger_\mu(\mbox{\boldmath$r$}
)-\Phi^2(\mbox{\boldmath$r$})\right]+\frac{\lambda}{\eta^2}\sum_{\mbox{\boldmath$r$}}\left\{\frac{1}{2}{\rm
Tr}[\Phi^2(\mbox{\boldmath$r$})]-\eta^2\right\}^2.
\label{higgs-action}
\end{equation}
Here we defined
$\Phi(\mbox{\boldmath$r$})=\sum_af(\mbox{\boldmath$r$})g_{a3}^A(\mbox{\boldmath$r$})\sigma_a$
and used the relationship
\begin{equation}
\int df\, f^2(\mbox{\boldmath$r$}) dV(\mbox{\boldmath$r$}) \sim \int
d\Phi(\mbox{\boldmath$r$})d\psi(\mbox{\boldmath$r$}),
\label{identity}
\end{equation}
where $\psi$ is the Euler angle parameterizing the little group of $\Phi(\mbox{\boldmath$r$})$.
Consequently, the integrand is independent of $\psi$.  To prove (\ref{identity}) we  use the
representation of the Haar measure by Euler angles and (\ref{g-adjoint}).

  Clearly, (\ref{higgs-action}) is just  a standard lattice Higgs action.  To show that SAG is indeed
equivalent to a Higgs theory we only need to investigate whether the presence of the term
$\tilde S$, defined in (\ref{stilde}), changes the nature of the action.
  
First of all, $\tilde S(U)$ is
a gauge invariant function of the gauge field, $U_\mu(\mbox{\boldmath$r$})$. Though it is
nonlocal, as we will show later, it becomes local in the relevant, $\eta\to \infty$, $g\to0$ limit. It is
also the generating functional of connected diagrams of an adjoint scalar field theory in a background
gauge field,
$U_\mu(\mbox{\boldmath$r$})$.  It must be composed of gauge invariant combinations of the
gauge field, Wilson loops.    $\tilde S(U)$ also has the property of local
$Z_2$ invariance, i.e. unlike the standard plaquette action it is invariant under the local replacement
$U_\mu(\mbox{\boldmath$r$})\to-U_\mu(\mbox{\boldmath$r$})$.  This is the property of a gauge
action with an adjoint representation trace of the plaquette product of gauge fields.  

To investigate the behavior of $\tilde S$ we first use a hopping parameter
expansion~\cite{montvay}. That implies  expanding the integrand  in
the hopping term,  integrating over the gauge group, and re-exponentiating
$\int dg(\mbox{\boldmath$r$}) Z^g_H(\lambda)$.  The hopping parameter expansion assumes
small $\eta$.  

The leading nontrivial term of this expansion is 
\[
\tilde
S(U)\simeq\frac{8
(f^2_{\rm
av})^4}{81}\sum_{\mbox{\boldmath$r$}\mu\nu}|U^P_{\mu\nu}(\mbox{\boldmath$r$})|^2=
\frac{8(f^2_{\rm
av})^4}{81}\sum_{\mbox{\boldmath$r$}\mu\nu}\left[U^{P,A}_{\mu\nu}(\mbox{\boldmath$r$})-1\right]
\]
where $U^{P,A}_{\mu\nu}(\mbox{\boldmath$r$})$ is the plaquette in adjoint representation and
\[
f^2_{\rm av} =\frac{\int_0^\infty df\,f^2 e^{-(\lambda/\eta^2)(f^2-\eta^2)^2}}{\int_0^\infty df\,
e^{-(\lambda/\eta^2)(f^2-\eta^2)^2}}\simeq \eta^2.
\]
 This adds a negative adjoint representation term to the usual plaquette term of the gauge action,
\[
\frac{1}{g^2(a)}U^P_{\mu\nu}(\mbox{\boldmath$r$})\to 
\frac{1}{g^2(a)}U^P_{\mu\nu}(\mbox{\boldmath$r$})-\Delta
\beta_AU^{P,A}_{\mu\nu}(\mbox{\boldmath$r$}),
\]
where in the leading order approximation $\Delta \beta_a=8(f^2_{\rm
av})^4/81.$

Provided the hopping parameter expansion is convergent, higher order terms can only have one of
three effects:
\begin{enumerate}
\item They can renormalize $\Delta\beta_A$,
\item Loop terms larger than the plaquette (nonlocal terms) may appear in the adjoint representation,
\item Characters in higher order integer isospin representations may also appear.
\end{enumerate}
Lattice gauge theories with  actions mixing the characters of plaquettes in fundamental and
adjoint representations have been studied thoroughly~\cite{mixed1}. The conclusion
of those investigations is that though such a modification of the action leads to new phase transition
lines
in the space of couplings. None of those transitions is deconfining. In the
continuum limit, when all couplings tend to zero, all of these theories are equivalent to a nonabelian
gauge theory with a modified coupling. 

In fact, there is no real need to evaluate all these terms, as we can predict the resulting effective
action using simple considerations.  The inclusion of the gauge fixing term
does not change the expectation value of any gauge invariant quantity of the pure gauge theory.
These quantities are dependent only on the gauge coupling, or alternatively, on the scale of the
running coupling, $g$.  The lattice coupling runs as
\[
\frac{1}{g^2(a)}-\frac{1}{g_0^2}\simeq 2\beta_G\log[\Lambda(a)/\Lambda_0],
\]
where $\Lambda(a)\sim 1/a$ is the lattice scale and $\beta_G=11/24\pi^2$; $\Lambda_0$ and $g_0$
refer to another, fixed scale.  

The action is also equivalent to an adjoint Higgs action with an effective
gauge coupling (the modification coming from the contribution of
$\tilde S$),
$g_{\rm eff}$.  The effective gauge coupling is such that the inclusion of the Higgs boson
changes its running into that of the non-gauge fixed coupling, namely
\[
\frac{1}{g_{\rm eff}^2} -\frac{1}{g_0^2}\simeq2(\beta_G+\beta_H)\log(\Lambda(a)/\Lambda_0),
\]
where  $\beta_H=1/24\pi^2$. 
 Physical quantities, depending on the gauge field only, must
be the same when the Higgs theory action
$S=S_G+\tilde S+S_H$ is used  as when the $S_G$ action is used alone.  As, at small $g(a)$ these
quantities are dependent only on the scale, $\Lambda(a)$, the scales in the two calculation must
coincide.  This leads to a relationship between the effective coupling and the original gauge
coupling 
\begin{equation}
g^2_{\rm eff}(a)\simeq g^2(a)\frac{\beta_G}{\beta_G+\beta_H}=\frac{11}{12}g^2(a),
\label{modification}
\end{equation}
valid at sufficiently small $g(a)$.

Let us return now to the discussion of the non-locality of $\tilde S$.\footnote{The author
is indebted to P. de Forcrand for calling his attention to this subject.} We can show that the
nonlocality goes away in the limit of
$\eta\to\infty$ and
$g\to0$ (continuum limit), not only in the limit of $\eta\to 0$. This is the very limit in which SAG
goes over into GAG and the lattice gauge theory goes over into the continuum theory.  First of all,
after rescaling
$\Phi\to \Phi\,\eta$, we have $ S_H\to \eta^2  {\tilde S}_H$, where $
{\tilde S}_H$ is independent of $\eta$.  As
\[
\tilde S = \log \int d\Phi\, e^{-S_H},
\]
 the large $\eta$ limit is equivalent to the semiclassical
limit as 
$\eta^2$ plays the role of $1/\hbar$.  In this limit only single loop terms contribute to 
$\tilde S$. In particular, all reference to the self-coupling of the Higgs field, $\lambda$, is of
$O(\eta^{-2})$, as it appears in multi-loop terms only.   As the external fields in
$\tilde S$ are gauge fields and the internal lines are Higgs propagators (there are no internal gauge
lines!) all the couplings of the one loop term are gauge couplings.  Then the calculation of $\tilde S$
is very similar to the calculation of the running gauge coupling using the background gauge field
method~\cite{peskin}. Thus, among the one loop contributions to the logarithm of the functional
determinant only diagrams with at most 4 background gauge fields are divergent.  As it is shown on
Ref.~\cite{peskin} the divergent contributions can be combined into the {\em local} term 
\[
\log\det \Delta=\int d^4x\left(\frac{1}{4}C(F_{\mu\nu}^a)^2+\mbox{ finite terms}\right),
\]
where $C$ is a logarithmically divergent constant, depending on the transformation properties of
the field in the loop.   When the running coupling is calculated~\cite{peskin} one must consider the
contribution of the dynamical gauge field, the fermion fields, and the ghost field, as well.  We only
need to calculate a scalar field loop, so the contribution is slightly different. We obtain
\[
\tilde S=-\frac{1}{2}\log(a\Lambda_{\rm lattice})\frac{1}{24\pi^2}\int
d^4x(F_{\mu\nu}^a)^2+\mbox{ terms finite at $a$=0},
\]
where the scale was chosen to be $\Lambda_{\rm lattice}$. 
As in the scaling limit, the gauge coupling in the pure gauge term scales with the lattice parameter
as
\[
\frac{1}{g^2}= \beta_G\log \left(\frac{1}{a\Lambda_{\rm lattice}}\right).
\]
$\tilde S$  combines  with $S_G$ to result in the replacement of $g^2$ by $g_{\rm eff}^2$, as
indicated in (\ref{modification}). At large $\eta$ and in the continuum limit the role of $\tilde S$ is
to renormalize the gauge coupling.  Then, in  the continuum limit, at large
$\eta$, the gauge fixed pure gauge theory is indeed equivalent to a Higgs theory.  It is clear that at
smaller values of $\eta$ or/and at larger gauge coupling  nonlocal effects are more
pronounced and $\tilde S$ is dependent on the parameters of the Higgs
potential.  Nonlocal effects are of $O(1/\log(a\Lambda)$ and of
$O(\eta^{-2})$.  We have also seen, however, that in the continuum limit at small $\eta$ these
theories are also equivalent.

 The equivalence of these lattice theories does not alone provide an obvious
relationship between monopole configurations found in them, as the monopoles themselves are not
gauge invariant objects.  To prove such an equivalence we need to examine  not only partition
functions but also Boltzmann factors.  

Assuming that the gauge coupling is small enough to apply continuum scaling,
 gauge configurations of the continuum theory and the lattice theory are essentially equivalent. 
To find classical solutions of the continuum theory we need to minimize the action in terms of the
fields.   The maximum of the Boltzmann factor should be calculated by
minimizing
$S_G+\tilde S+S_H$ in terms of the gauge fields and gauge transformations, as well.  This
minimization is  equivalent to the minimization in terms of the Higgs field and the gauge field.  In
the continuum limit this minimization should lead to a combination of configurations containing a
variable number of  finite energy classical solutions, monopoles ~\cite{thooft4}~\cite{polyakov}.
 Monopoles are identified by the  radial
dependence of the magnitude of the Higgs field. The magnitude of the Higgs field itself is gauge
invariant.  The final step in  identifying the monopoles with those of GAG is the diagonalization of 
the Higgs field with a gauge transformation.  

The conclusion of the above considerations is that
\begin{enumerate}
\item
The appearance of monopoles in  gauge fixed pure gauge theories follows from their existence in
Higgs theories.
\item
The analytic form, interaction, condensation, etc. of monopoles should be very similar to those
in Higgs theory.
\item
The mechanism of confinement, provided monopoles are responsible for it, is the same in pure gauge
and Higgs theories.
\end{enumerate}
Before turning to vortices let us apply our method to calculate  the
distribution of a single monopole in Abelian projected GAG.  To do that
we take the  the 't Hooft-Polyakov monopole and diagonalize the Higgs boson by a gauge
transformation.  The
 gauge and Higgs fields are given by
\begin{eqnarray}
\mbox{\boldmath$A$}(\mbox{\boldmath$r$})&=& \mbox{\boldmath$r$}\times
\mbox{\boldmath$\sigma$}
\, F(r),
\nonumber\\ \Phi(\mbox{\boldmath$r$})&=& \mbox{\boldmath$r$}\cdot
\mbox{\boldmath$\sigma$}\, W(r),
\label{ansatz1}
\end{eqnarray}
where $F(r)$ and $W(r)$ are the radial wave functions. 

Then the abelian gauge transformation is a rotation around the
$z$ axis by an angle $-\phi$, followed by a rotation around the $y$ axis by an angle $-\theta$, where
$\phi$ and $\theta$ are polar coordinates. Applying the same  gauge transformation to the
gauge field   and projecting to the abelian component we obtain (we set the gauge charge equal to 1)
\[
\frac{1}{2}{\rm Tr}[\sigma_z(UA_\mu U^\dagger+iU\partial_\mu U^\dagger)]=
\frac{z}{r}\partial_\mu\phi.
\]
  The magnetic field obtained from this vector potential is that of a single magnetic charge at the
origin 
\[
\mbox{\boldmath$B$}=\frac{1}{2}\frac{\mbox{\boldmath$\hat r$}}{r^2}+\mbox{ Dirac string}.
\]

It is worth noting that the abelian projection of gauge field (\ref{ansatz1}) {\em without abelian
gauge fixing} leads to
\[
A_\mu(\mbox{\boldmath$r$})=\partial_\mu \phi\,r^2F(r),
\]
that is a more like a vector potential  for a magnetic flux than for a monopole. 
 
\section{ Central gauges and Higgs theories}
Now we turn  to center gauges and vortices.    Lattice studies have shown the relevance of $Z_2$
vortices  in pure  lattice
$SU(2)$ gauge theory.   These studies use two alternative gauges,
  the Maximal Center gauge (MCG)~\cite{vortex1} and the Laplacian Center
gauge (LCG)~\cite{vortex2}~\cite{vortex3}. The gauge transformations are followed by center
projection.  

MCG is defined through the maximization of the following functional (we use notations,
identical to the ones used for the discussion of Abelian gauges) over gauge transformation $V(x)$:
\begin{equation}
F^C(U)=\frac{1}{4}\sum_{x\mu}|{\rm
Tr}U_\mu^V(\mbox{\boldmath$r$})|^2=\frac{1}{6}\sum_{\mbox{\boldmath$r$}\mu}{\rm
Tr}\left[\sum_i^3\sigma_iU_\mu^V(\mbox{\boldmath$r$})\sigma_iU_\mu^{V\dagger}(\mbox{\boldmath$r$}
)\right].
\label{center1}
\end{equation}
$F^C(U)$ can be rewritten in terms of adjoint representation quantities as
\begin{equation}
F^C(U)=\frac{1}{3}\sum_{\mbox{\boldmath$r$}\mu}
\sum_{iab}^3g^A_{ai}(\mbox{\boldmath$r$})U_\mu^{A,ab}(\mbox{\boldmath$r$})
g^A_{bi}(\mbox{\boldmath$r$}+\mbox{\boldmath$\mu$}).
\label{adjoint2}
\end{equation}
If one writes the gauge field in the canonical form
\[
U_\mu^g(\mbox{\boldmath$r$})=a_0+i\mbox{\boldmath$a\cdot\sigma$},
\]
where $a_0^2+\mbox{\boldmath$a$}^2=1$, then $F^C(U)=\sum a_0^2$.  The maximization of
$\sum a_0^2$ results in a gauge field as close to the center of the gauge group ($a_0=\pm1$) as
possible.  

In fact,
it is sufficient to keep only two terms of the sum over $i$ in (\ref{center1}) and (\ref{adjoint2}), as
the adjoint representation gauge transformation is defined uniquely by two of the orthogonal vectors
$g_{a1}^A$ and
$g_{a2}^A$.  This is used in LCG, which relaxes the normalization and orthogonality conditions on
the vectors
$g_{ai}^A$.  Vortices are found  at points where the
two vectors are parallel. These points form lines in three dimensional space.  At these points the
gauge transformation is singular.  In what follows we will restrict ourselves to such two term gauge
fixing functionals, though everything what we do can easily be generalized to the three term case, 
such as the original MCG.

MCG and LCG can again be extended into a Generalized Center Gauge
(GCG),  which also fixes the gauge, up to the center of the
$SU(2)$ gauge group, $Z_2$.  A convenient form for the gauge fixing functional is 
\begin{equation}
Z_C(U^V)= \int d\Phi^1(\mbox{\boldmath$r$})\,\int_0^\infty
df(\mbox{\boldmath$r$})\,\rho^2(x)\exp\left\{-
\tilde F_{\rm
gf}(U^g,f)-S_H(U,\Phi^1)-S_{\rm mix}(V,f,\Phi^1)\right\}
\label{intermediate2}
\end{equation}
where $\tilde F_{\rm gf}(U^V,f)$ and $S_H(U,\Phi^1)$ were defined in (\ref{higgs-action0})
and (\ref{higgs-action}), respectively, while the `mixing term' is given by
\[
S_{\rm mix}(V,f,\Phi^1)=\frac{\tilde\lambda}{\eta^2}
\sum_{\mbox{\boldmath$r$}}[
f(\mbox{\boldmath$r$})g_{a3}^A(\mbox{\boldmath$r$})\Phi_a^1(\mbox{\boldmath$r$})-c\eta^2]^2,
\]
where $c$ is the cosine of the asymptotic angle between $g^A_{a3}$ and $\Phi_a$.
When we maximize
(\ref{intermediate2}) in terms of
$V(\mbox{\boldmath$r$})$ and $f(\mbox{\boldmath$r$})$ we obtain an abelian gauge. The
exponent will be dominated by the  configuration of the Higgs field, gauge transformation, and
$f(\mbox{\boldmath$r$})$ that maximizes the exponent.  In contrast to the Abelian gauge fixing of
the previous section this limit fixes the abelian gauge group, as well.  The Abelian subgroup is fixed 
by a rotation around the 3rd axis (the little group of
$g_{a3}^A(\mbox{\boldmath$r$})$), such that
$\Phi_2^1(\mbox{\boldmath$r$})=0$. Vortices  appear on lines in the 3D space where
$\Phi_a^1(\mbox{\boldmath$r$})$ is parallel with
$g_{a3}^A(\mbox{\boldmath$r$})$ and where such a gauge fixing is impossible.  Alternatively, in a
slightly more symmetric manner, we could also define the gauge transformation by rotating  the sum
of the normalized Higgses parallel to the $z$ axis and rotating the difference into the $(xz)$-plane. 

When $c=0$ and $\lambda,\ \tilde\lambda\to\infty$ GCG reduces to  MCG (or rather an analogous 2
term gauge), as
$\Phi^1_a(\mbox{\boldmath$r$})$ is kept normalized to $\eta$ and orthogonal to
$g_{a3}^A(\mbox{\boldmath$r$})$.  When
$\lambda_1=\lambda_2=0$ GCG reduces to LCG. 

Next we introduce the smeared center gauge  (SCG) such that the gauge fixing term in the
Lagrangian is defined as
\[
e^{-S_{\rm gf}}= Z_C(U^g)e^{-\bar S},
\label{scg}
\]
where
\[
e^{\bar S}=\int dV'(\mbox{\boldmath$r$})Z_C(U^{V'})
\]
In a manner similar to our discussion of Abelian gauges we can show that on large lattices 
SCG is very close to GCG and we can also show that SCG is completely equivalent to a lattice Higgs
theory with two adjoint Higgses. 
This follows because after a gauge transformation one can introduce the second Higgs field as 
\[
\Phi^2_a(\mbox{\boldmath$r$})=V(\mbox{\boldmath$r$})g_{a3}^A(\mbox{\boldmath$r$})
\]
and prove that the action can be written as $S=S_G+S_{\rm gf}=S_G+\bar S+S^2_H$, where 
$S_H^2$ is a Lagrangian for two adjoint Higgses.
$S_G+\bar S$ is again a modified pure gauge action. The modified gauge coupling should be
equal to the standard pure gauge coupling minus the contribution of two adjoint Higgs bosons, as
$S$ is equivalent to a gauge fixed version of the pure gauge theory. Thus the relationship between
the gauge couplings is $g_{\rm eff}^2(a)\simeq (11/13)g^2(a).$

The Boltzmann factor of
the GCG must be dominated by vortex configurations,  if the Boltzmann factor of the Higgs
theory is.  To make the connection between GCG and SCG distributions we need to  fix the vortex
gauge such that one of the Higgs bosons is rotated parallel to the 3rd  axis in isospace while the
second Higgs boson is rotated into the
$1-3$ plane.  Thus, the `thick' vortices appearing in GCG have the same location and radial
dependence as the vortices of the Higgs theory. 

To illustrate the  equivalence of GCG and SCG we calculate  the shape of a single
vortex in center gauge fixed pure gauge theory, by making use of a classical solution of the Higgs
theory with two adjoint Higgs bosons~\cite{suranyi}. The solution is of the form 
\begin{eqnarray}
\Phi^{1,2}(\mbox{\boldmath$r$})&=&U(\phi)
[\chi_d(\rho)\sigma_3\pm\chi_\bot(\rho)\sigma_1]U^\dagger(\phi),
\nonumber \\
A_\mu(\mbox{\boldmath$r$})&=& a(\rho) \partial_\mu \phi\,\sigma_3,
\label{vortex-solution}
\end{eqnarray}
where the boundary conditions require that $\chi_\bot(0)=a(0)=0$ and $a(\infty)=1$, the
topological charge.  $\rho$ and $\phi$ are cylindrical coordinates, along with $z$ and
\[
U(\phi)=e^{i\phi\sigma_3/2}.
\]
The normalized components of the Higgs field
also satisfy boundary conditions at infinity: $\chi_\bot(\infty)=\sqrt{(1-c)/2}$,
$\chi_d(\infty)=\sqrt{(1+c)/2}$.
$c$ is the angle between the two Higgs bosons at infinity.   
(\ref{vortex-solution})
satisfies the constraint that the two Higgs fields become parallel at the location of the vortex, along
the $z$ axis. The $z$ axis is also the locus of the singularity of $U(\phi)$. 

To predict the form of a thick vortex on the lattice, in pure gauge theory, we need to bring the
Higgs bosons to a standard form. This is the form in which the fields have only components in
the
$(xz)$ plane and the sum of the Higgses is diagonal.  To get to that form we need to use the singular
gauge transformation
$U^\dagger(\phi)$.  As $U$ commutes with the gauge field of (\ref{vortex-solution})  the only
change of the gauge field will be the addition of the inhomogeneous term, $iU^\dagger\partial_\mu
U$. This will result in the following change in the gauge field (the gauge charge has been set equal to
unity):
\[
A_\mu(\mbox{\boldmath$r$})\to A_\mu(\mbox{\boldmath$r$})+iU^\dagger\partial_\mu
U=[a(\rho)-1] \partial_\mu
\phi\,\sigma_3.
\]
The gauge field remains diagonal after the gauge transformation. It is singular on the $z$-axis and it
goes to zero exponentially at
$\rho=\sqrt{x^2+y^2}\to\infty$. 

To understand what happens on the lattice and after center projection
we rewrite the Higgs fields, the vector potential, and the gauge transformation into a form
appropriate for the  lattice.  The Higgs field (\ref{vortex-solution}) is already in such a form.
The lattice gauge field is a unitary matrix. The gauge field in (\ref{vortex-solution}) can be
conveniently written as
\begin{equation}
U_\mu(\mbox{\boldmath$r$})=\exp\left\{
\frac{i}{2} a(\rho(\mbox{\boldmath$r$}))[\phi(\mbox{\boldmath$r$}+
\mbox{\boldmath$\mu$})-\phi(\mbox{\boldmath$r$})]\sigma_3
\right\},
\label{lattice-gauge}
\end{equation}
where $\rho(\mbox{\boldmath$r$})$ and $\phi(\mbox{\boldmath$r$})$ are the cylindrical
coordinates at lattice points. The singular gauge transformation that brings the Higgs field into the
$(xz)$ plane is 
\[
V(\mbox{\boldmath$r$})=\exp\left\{\frac{i}{2}\phi(\mbox{\boldmath$r$})\right\}.
\]
Thus, the gauge transformed gauge field is 
\[
U_\mu^V(\mbox{\boldmath$r$})=V(\mbox{\boldmath$r$})U_\mu(\mbox{\boldmath$r$})
V^\dagger(\mbox{\boldmath$r$}+
\mbox{\boldmath$\mu$})=
\exp\left\{
\frac{i}{2}
[a(\rho(\mbox{\boldmath$r$}))-1][\phi(\mbox{\boldmath$r$}+\mbox{\boldmath$\mu$})-\phi(
\mbox{\boldmath$r$})]\sigma_3\right\}.
\]

Let us examine the gauge field near the $z$ axis. At $\rho\sim 0$ the radial component  of the
original gauge field is negligible,
$a(\rho)\sim \rho^2$, and the new gauge field has the form of a pure gauge
transformation.  When one traverses around the $z$ axis one needs to define a surface in the three
dimensional space, with a boundary at the $z$ axis, such that on this surface the value of $\phi(x)$
jumps by $2\pi$.  This surface can be chosen arbitrarily, so we choose the  half-$(xz)$ plane
containing the negative half of the $x$ axis.  If the gauge coupling small, the argument of the
gauge field $U_\mu(x)$ is small and at central projection projects to the identity matrix.  When we
intersect the half plane, however, 
\[
U_y(\mbox{\boldmath$r$})\simeq e^{\frac{i}{2}[\pi-(-\pi)]}=-1,
\]
and the gauge field projects to the non-trivial element of the center, -1. This sheet of negative
elements of the center ends in the $z$-axis, where a $Z_2$ vortex appears.  Thus, we showed that in
the center gauge, after center projection a $Z_n$ vortex appears wherever a vortex appears in the
Higgs theory. 
 At the same
time we have given a representation for the thick vortex, obtained after transformation to center
gauge, but before center projection.   

Strictly speaking, a secondary vortex also appears at some distance from the $z$ axis.  The location
of that vortex is determined by the equation
\[
a(\rho)-1\simeq \frac{1}{2},
\]
which ensures that the jump of the argument of $U_\mu$ is smaller than $\pi/2$.
At large $\rho$ the radial component of the gauge field  $a(\rho)\sim e^{-m_A\rho}$, where the
auxiliary gauge mass,
$m_A=g(a)\eta$. Then  the secondary vortex will appear at $\rho\sim1/m_A$. If, in the continuum
limit, 
$g(a)$ is sufficiently small then this vortex will be far away from the $z$ axis, not affecting the
confinement picture.  The location of the primary vortex is gauge invariant and of the secondary
vortex depends on the gauge.  On the contrary, 
 if we perform the center
projection without transforming (\ref{vortex-solution}) to center gauge, 
then the $Z_2$ vortex does
not appear at the
$z$ axis, we would only have the above described secondary vortex and the simple vortex
condensation picture would not be recognizable. 

\section{Conclusion}

We have shown that pure gauge theories in Abelian or
Center gauges are equivalent to Higgs theories with one or more adjoint representation Higgs
bosons. The equivalence extends to gauge configurations.  Monopoles and vortices and their
interactions, condensation, etc. in  gauge fixed pure gauge   theories can be studied analytically by
investigating similar objects in Higgs theories, where they appear as classical solutions.

We encounter a problem when we apply our method of connecting pure gauge theories with Higgs
theories in the limiting cases of Maximal Abelian and Maximal Center gauges.  Then our
construction leads to gauged nonlinear sigma
models.  As far as we know, the relevant gauged nonlinear sigma models do not have classical
solutions in 3+1 dimensions.  The reason is simple:  Higgs fields in nonlinear sigma models cannot
vanish.  In gauged linear sigma models the location of the zeros marks the center of monopoles or
vortices.  This is, of course, in agreement with the fact that monopoles and vortices do not appear as
singularities of the gauge transformation in MAG and MCG either. It would be interesting to
investigate, using  nonlinear sigma models, why the projected theories still show many of the
characteristics found in General Abelian  and General Center gauges. 

 The link between gauge fixed pure gauge theories and Higgs theories can be
employed to investigate important questions concerning confinement. One example that comes to
mind is  the recently uncovered intriguing relationship between monopole and vortex
condensations~\cite{deldebbio}~\cite{ambjorn}.  Simulations show that monopoles do
not form a Coulomb gas as was originally assumed, but rather they line up to form vortices,
thereby making vortices more fundamental for the mechanism of confinement. 

Another important question is  why multiple non-abelian vortex
configurations appear at all.   `t Hooft's argued~\cite{thooft5} that in a $SU(2)/Z_2$ theory the
homotopy group has only one non-trivial class, so vortices should coalesce into a single
vortex or no vortex at all.  The situation is somewhat different in $U(1)$ gauge theory where the
homotopy group is
$Z$, and infinitely many kinds of vortices exist.  Lattice evidence supports, however, the
existence of multi-vortex configurations.  It is conceivable that studying  interactions of vortices
 in Higgs theories one will be able to answer this important question, as well. 

The author thanks  the U.S. Department of Energy for partial support through grant
\#DE FG02-84ER-40153.

\end{document}